\documentclass[twocolumn,showpacs,pre,preprintnumbers,amsmath,amssymb,aps,floatfix]{revtex4}
\usepackage{graphicx}
\usepackage{dcolumn}
\usepackage{bm}
\begin{document}
\title{Transmission of Information between Complex Networks:
 $1/f$-Resonance.}
\newcommand{\e}{{\rm e}}
\newcommand{\veps}{\varepsilon}
 \newcommand{\Sss}{\scriptscriptstyle}
\newcommand{\Ss}{\scriptstyle}
\newcommand{\D}{\dysplaystyle}
\newcommand{\T}{\textstyle}
\newcommand{\lgl}{\langle}
 \newcommand{\rgl}{\rangle}
\newcommand{\Vh}[1]{\hat{#1}}
\newcommand{\Aa}{A^1_{\epsilon}}
\newcommand{\Ab}{A^{\epsilon}_L}
\newcommand{\Ae}{A_{\epsilon}}
\newcommand{\finn}[1]{\phi^{\pm}_{#1}}
\newcommand{\ea}{e^{-|\alpha|^2}}
\newcommand{\eb}{\frac{e^{-|\alpha|^2} |\alpha|^{2 n}}{n!}}
\newcommand{\ebbb}{\frac{e^{-3|\alpha|^2} |\alpha|^{2 (l+n+m)}}{l!m!n!}}
\newcommand{\ass}{\alpha}
\newcommand{\as}{\alpha^*}
\newcommand{\fb}{\bar{f}}
\newcommand{\gb}{\bar{g}}
\newcommand{\la}{\lambda}
 \newcommand{\sz}{\hat{s}_{z}}
\newcommand{\sy}{\hat{s}_y}
\newcommand{\sx}{\hat{s}_x}
\newcommand{\sio}{\hat{\sigma}_0}
\newcommand{\six}{\hat{\sigma}_x}
\newcommand{\siz}{\hat{\sigma}_{z}}
\newcommand{\siy}{\hat{\sigma}_y}
\newcommand{\vhsig}{\vec{\hat{\sigma}}}
\newcommand{\hsig}{\hat{\sigma}}
\newcommand{\hH}{\hat{H}}
\newcommand{\hU}{\hat{U}}
\newcommand{\hA}{\hat{A}}
\newcommand{\hB}{\hat{B}}
\newcommand{\hC}{\hat{C}}
\newcommand{\hD}{\hat{D}}
\newcommand{\hV}{\hat{V}}
\newcommand{\hW}{\hat{W}}
\newcommand{\hK}{\hat{K}}
\newcommand{\hX}{\hat{X}}
\newcommand{\hM}{\hat{M}}
\newcommand{\hN}{\hat{N}}
\newcommand{\te}{\theta}
\newcommand{\vze}{\vec{\zeta}}
\newcommand{\vet}{\vec{\eta}}
\newcommand{\vx}{\vec{\xi}}
\newcommand{\vc}{\vec{\chi}}
\newcommand{\hro}{\hat{\rho}}
\newcommand{\vro}{\vec{\rho}}
\newcommand{\hR}{\hat{R}}
\newcommand{\half}{\frac{1}{2}}
\renewcommand{\d}{{\rm d}}
\renewcommand{\top }{ t^{\prime } }
\newcommand{\oz}{{(0)}}
\newcommand{\sint}{{\rm si}}
\newcommand{\cint}{{\rm ci}}
\newcommand{\de}{\delta}
\newcommand{\ep}{\varepsilon}
\newcommand{\De}{\Delta}
\newcommand{\eps}{\varepsilon}
\newcommand{\si}{\hat{\sigma}}
\newcommand{\om}{\omega}
\newcommand{\tr}{{\rm tr}}
\newcommand{\ha}{\hat{a}}
\newcommand{\gam}{\gamma ^{(0)}}
\newcommand{\pe}{\prime}
\newcommand{\BEQ}{\begin{equation}}
\newcommand{\EEQ}{\end{equation}}
\newcommand{\BEA}{\begin{eqnarray}}
\newcommand{\EEA}{\end{eqnarray}}
\newcommand{\sph}{spin-$\frac{1}{2}$ }
\newcommand{\ad}{\hat{a}^{\dagger}}
\newcommand{\add}{\hat{a}}
\newcommand{\spp}{\hat{\sigma}_+}
\newcommand{\smm}{\hat{\sigma}_-}
\newcommand{\fin}[1]{|\phi^{\pm}_{#1}\rangle}
\newcommand{\finp}[1]{|\phi^{+}_{#1}\rangle}
\newcommand{\finm}[1]{|\phi^{-}_{#1}\rangle}
\newcommand{\lfin}[1]{\langle \phi^{\pm}_{#1}|}
\newcommand{\lfinp}[1]{\langle \phi^{+}_{#1}|}
\newcommand{\lfinm}[1]{\langle \phi^{-}_{#1}|}
\newcommand{\lfinn}[1]{\langle\phi^{\pm}_{#1}|}
\newcommand{\z}{\cal{Z}}
\newcommand{\RI}{\hat{{\cal{R}}}_{0}}
\newcommand{\Rt}{\hat{{\cal{R}}}_{\tau}}
\newcommand{\nn}{\nonumber}


\author{Gerardo Aquino$^1$}
\author{Mauro Bologna $^{2}$ }
\author{Bruce J. West$^3$}
\author{Paolo Grigolini$^{4}$}
\affiliation{$^1$ Faculty of Natural Sciences, Imperial College London, SW7 2AZ London, UK}
\affiliation{$^2$ Instituto de Alta Investigaci\'on, Universidad
de Tarapac\'{a}-Casilla 7-D Arica, Chile}
\affiliation{$^{3}$ Physics Department, Duke University, North Carolina, USA.}
 \affiliation{$^4$Center
for Nonlinear Science, University of North Texas, P.O. Box 311427,
Denton, Texas 76203-1427, USA}
\date{\today}
\begin{abstract}
We study the transport of information between two complex  networks with similar properties.
Both networks generate non-Poisson renewal fluctuations with a power-law spectrum  $1/f^{3-\mu}$, the case $\mu=2$ corresponding to  ideal $1/f$-noise.  We denote by $\mu_S$ and $\mu_P$ the power-law
indexes of the  network ``system'' of interest $S$ and the perturbing network $P$
respectively.
By adopting a generalized fluctuation-dissipation theorem (FDT) we show that the ideal condition of $1/f$-noise for both networks corresponds to maximal information transport.
 We prove that to make the network $S$ respond
 when $\mu_S < 2$ we have to set the condition $\mu_P <
2$. In the latter case, if $\mu_P<\mu_S$, the system $S$ inherits the relaxation properties of the
 perturbing network. In the case where $\mu_P > 2$,  no response and no information
transmission occurs in the long-time limit. 
 We consider two possible generalizations of the fluctuation dissipation theorem
and show that both lead to maximal information transport in the condition of $1/f$-noise.

     \end{abstract}

\pacs{05.40.Fb, 05.40.-a, 02.50.-r,,82.20.Uv}

\maketitle

\section{Introduction}\label{Newintroduction}
 Linear Response Theory (LRT) \cite{kubo} is one of the basic ways of obtaining information from fluctuations  in  non equilibrium statistical physics \cite{crisanti,vulpianone}
that is currently adopted to address new phenomena such as glassy systems \cite{crisanti} and granular matter \cite{vulpianone}. An even more challenging issue is the application of  LRT to complex processes such as physiological processes and especially the understanding of brain dynamics. 

It is becoming more widely accepted that the brain operates at criticality \cite{chialvocriticality,werner} and that the critical condition of a phase
transition has manifestations that extend beyond the  conventional condition
of temperature driven systems \cite{timephasetransition}. Frantsuzov \emph{et al.} \cite{notredame} adopted a model of cooperatively interacting units to propose a solution  to the long-standing mystery of the origin of the power-law distribution  of the blinking times in colloidal quantum dot fluorescence. On the other hand, this form 
of intermittence is characterized by the condition of renewal aging \cite{brokman}
and by the consequent ergodicity breakdown \cite{ergodicitybreakdown} that makes it impossible to use 
 conventional LRT. We refer to these systems as \emph{complex networks}. 
It is now very well understood \cite{brucephysrep} that the breakdown of the ergodic condition is caused by the occurrence of \emph{crucial events}. The important fact that cooperation-induced phase transition turns a regular network into a complex network, namely, a network
with temporal complexity and crucial events, has been proved in the recent work of Ref. \cite{gosia}.  These events are renewal, namely, the time interval between two consecutive events
does not have any relation whatsoever with the earlier or later time intervals between  two consecutive events. Yet, if a Gibbs ensemble is suitably \emph{prepared} initially, namely, in 
all the networks of the ensemble an event occurs at the time origin, then the rate of event production turns out to be time dependent rather than constant as in the ordinary Poisson case. The authors of Refs. \cite{cascade,liquid} applied this theoretical perspective to the liquid crystal dynamics and experimentally realized a true cascade of renewal events.  

The time interval between two consecutive crucial events is given by a waiting times probability density function (pdf) $\psi(\tau)$ with the following asymptotic form
\begin{equation} \label{specialform}
\psi(\tau)  = (\mu-1) \frac{T^{\mu-1}}{\left(\tau + T\right)^{\mu-1}}\end{equation}
and
 the power-law index $\mu$ fulfilling the  inequality
\begin{equation}
\label{crucial}
1 < \mu < 3. 
\end{equation}

It is important to explain the origin of the special form 
of Eq. (\ref{specialform}). First of all we want to stress that according to a  point of view in the field of complexity, only the asymptotic time behavior matters, namely, $\psi(\tau) \propto 1/\tau^{\mu}$. The adoption of this widely shared point of view, as we shall see hereby, would prevent us from establishing 
a correct accordance with the experiments on the response of complex networks to external perturbations. Thus, the choice of Eq. (\ref{specialform}) is dictated by the need for defining a border between the asymptotic time regime $(\tau \gg T)$
and the microscopic time regime $(\tau \leq T)$. Given the neurophysiology interest of this paper  and especially  the focus on brain dynamics, we refer the interested reader to the work of Ref. \cite{fechner}, where the waiting times pdf of Eq. (\ref{specialform}) is obtained by means of a Fechner transformation \cite{fechner}   from the conventional Poisson distribution. 

The rate of cascade of renewal events tends to a vanishing value as $1/t^{2-\mu}$ when $\mu < 2$ and to a constant value as $1/t^{\mu-2}$ when $\mu > 2$. It is evident that in both cases the time duration of the out-of-equilibrium condition is infinite, thereby raising the challenging task of going beyond  
conventional LRT to describe the dynamics.

Conventional LRT is given by the following expression:
\begin{equation}
\label{LRT}
\sigma(t)=\left<\xi_{S}(t) \right> = \epsilon \int_{0}^{t} ds \chi(t,s) \xi_{P}(s),
\end{equation}
where $\xi_{S}(t)$ is the fluctuation produced by the network of interest $S$.
The symbol $\left<\xi_{S}(t) \right>$ denotes the Gibbs average over the fluctuations. In the absence of perturbation this average is assumed to vanish.
The variable $\xi_{P}(t)$ denotes the time dependent perturbation and $\epsilon$ its intensity. LRT predicts the response of $S$ on the basis of the unperturbed correlation function
of $\xi_{S}$. In fact the function $\chi(t,s)$, called the linear response function,
is related to the correlation function of the fluctuation $\xi_{S}$, whose quadratic mean value is assumed for simplicity to be normalized to  unity,
\begin{equation}
\Psi_{S}(t,s) \equiv \left<\xi_{S}(t) \xi_{S}(s)\right>\end{equation}
by the following expression
\begin{equation}
\chi(t,s) = \frac{d}{ds} \Psi_{S}(t,s). 
\end{equation}
Note that the traditional LRT refers to the stationary case
\begin{equation}
\Psi_{S}(t,s) = \Psi_{S}(t-s),
\end{equation}
and as a consequence
\begin{equation}
\frac{d}{ds} \Psi_{S}(t,s)  = - \frac{d}{dt} \Psi_{S}(t,s).
\end{equation}
This condition is not fulfilled by complex networks. For these  latter networks
the choices of linear response functions
\begin{equation} \label{phenomenological}
\chi(t,s) = \frac{d}{ds} \Psi_{S}(t,s)
\end{equation}
and
\begin{equation} \label{dynamical}
\chi(t,s) = - \frac{d}{dt} \Psi_{S}(t,s)
\end{equation}
are not equivalent. 

The authors of Refs. \cite{barbi,ultimoallegro,primogerardo,budinone}
have discussed the foundation of both choices and have established 
that the new form of LRT is determined by the physical way through which
perturbation determines a bias.  For the sake of simplicity  these authors have made the assumption that $\xi_{S}(t)$ is a dichotomous signal. Using the jargon of 
turbulence theory they called the time intervals between two consecutive crucial events  \emph{laminar regions} . At the moment of a crucial event occurrence, unperturbed dynamics are realized by the random selection of either the positive, $\xi_S = 1$, or the negative value, $\xi_S = -1$. In other words, they assume that the occurrence of a crucial event generates the tossing of a fair coin which determines the sign of the next laminar  region. Consequently, the external perturbation can generate a bias in two different ways. The former way rests on
affecting the fairness of the coin tossing process. If $\xi_{P}(t) > 0$  $(\xi_P(t) < 0)$ the choice of the positive (negative) sign is more probable than the choice of the negative (positive) sign. This prescription leads to the choice of Eq. (\ref{phenomenological}) and is denoted as \emph{phenomenological} LRT. 

The experiments done by the authors of Refs. \cite{liquid,cascade} show that  nature prefers Eq. (\ref{dynamical}), the dynamic LRT. What is the theoretical argument in favor
of the dynamical theory? To afford a convincing answer to this important question, let us go back to the special form of Eq. (\ref{specialform}). We note that we do not know the Hamiltonian of our complex network, and we do not even know if a satisfactory discussion of the complex dynamics can be made using a Hamiltonian formalism. Let us assume that Eq. (\ref{specialform}) is a reliable representation for the distribution length of the laminar region. In this case, a reasonable conjecture is that the external perturbation affects either $\mu$ or $T$, or both parameters defining the form of Eq. (\ref{specialform}). We know that $\mu$ defines the network's complexity and emerges from the cooperative interaction among interacting units. A weak external perturbation is not expected to change the network's complexity. It is therefore reasonable to assume that the external perturbation affects $T$, by enlarging (reducing) its value if $\xi_P$ and $\xi_S$ have the same (opposite) sign. 
This assumption leads to the choice of Eq. (\ref{dynamical}), as shown  earlier  \cite{barbi,ultimoallegro,primogerardo,budinone}.

It is important to notice that the response of a complex network of the same nature as the one discussed in this paper has been studied by many authors \cite{sokolov1,sokolov2,heinsalu1,heinsalu2,weron1,sokolovall, sushin, magiaro,kang,australians}. These authors did not establish a connection between their results and the LRT
of Refs. \cite{barbi,ultimoallegro,primogerardo,budinone}
and in some cases they made the misleading conjecture that their results establish the ``death of LRT''.  Actually, these theoretical treatments are asymptotic in time and the only possible connection with  LRT is through the adoption of the phenomenological theory of Eq. (\ref{phenomenological}), as the readers can establish by a careful reading of Ref. \cite{sokolov1}.

We are now in a position to define the main purpose of this paper. We draw the attention of the readers to  the recent results of Ref.\cite{pinknoise}. This paper addresses the important issue of studying the response of a complex network to a complex external perturbation with the surprising result that a complex network
does not respond to stimuli that are not complex, i.e. that have a stationary Fourier spectrum. It is important to stress that 
Ref. \cite{pinknoise} focuses on the correlation between $\xi_{S}(t)$ and $\xi_{P}(t)$ in the long time limit. This is an ideal condition that has the effect of restraining the definition of complexity to the networks with $\mu \leq 2$. In fact, in the long-time limit a network with $\mu > 2$ reaches the normal condition of a constant rate of event production, thereby recovering the ordinary Poisson condition. 
The condition $\mu = 2$ is of fundamental importance for  brain function. In fact, recent work \cite{menicucci1,menicucci2} established that the brain works with $\mu= 2$, which, in turn, is known \cite{mirko} to correspond to making the brain action become the source of ideal $1/f$ noise.  The results of Ref. \cite{pinknoise}  may  therefore have  important applications to design the most convenient stimuli to drive complex networks, and especially  brain dynamics, via what was defined as  ``complexity management''  \cite{pinknoise}.  However, an apparent weakness of Ref. \cite{pinknoise} is that these  results are derived from the adoption of the phenomenological LRT, thereby raising the doubt that the complex networks, which have been proven to obey the dynamical LRT \cite{cascade,liquid}, may not obey the principle of {\it complexity management} (CM) established in Ref. \cite{pinknoise}. Herein we  prove that the more realistic dynamical LRT generates CM. In addition to this main  purpose,  the present paper affords technical details
on the theory developed in Ref. \cite{pinknoise} that, due to space limitations, were not conveniently illustrated.


\section{A  FDT for non-ergodic renewal networks: phenomenological and dynamical approach}\label{nonergodicfdt}
The authors of Ref. \cite{ultimoallegro,primogerardo} 
discovered a form of FDT that applies to networks dominated
by non-Poisson renewal events. In the stationary case this FDT
becomes indistinguishable from the ordinary theoretical prediction
\cite{kubo}.

Herein we investigate the consequences of the adoption of
either the `` phenomenological'' choice  Eq. (8) or the ``dynamical'' choice  (9), in 
 the special case where both $\xi_{P}(t)$
and $\xi_{S}(t)$, are event dominated processes and show that
 the transmission of
information from $P$ to $S$ is determined by the dialogue between
the critical events of $\xi_S(t)$ and the critical events of
$\xi_P(t)$. Specifically, this discussion is devoted to studying
the transport of information from $P$ to $S$, using both forms of generalized FDT (gFDT).
 Note that there is no limitation on the form of
$\xi_{P}$, provided that the coupling is weak enough as to be
compatible with the emergence of the linear response form of Eq.
(3). Thus, in this paper we imagine that $P$ generates
a fluctuating signal $\xi_P(t)$ and that for any signal $\xi_P(t)$
there exists a response $\xi_S(t)$. We have a single composite
network $S + P$ and consequently a single signal $\xi_{S}(t)$. To
discuss the problem of the transmission of information from $P$ to
$S$, it is convenient to imagine the ideal case of a Gibbs
ensemble of networks $S+P$. For simplicity we take both signals
$\xi_S(t)$ and $\xi_P(t)$ to be dichotomous and fluctuating between values
$\pm 1$

It is important to remark that Eq. (\ref{LRT}) for the
response of the ``system'' network $S$ to the ``perturbing'' network $P$ is valid when the network
is prepared at time $t=0$ and the interaction with the perturbation $P$
is turned on at the same time. 
Nothing is said about $\xi_{P}(t)$ because Eq. (3) is based on the assumption that for each perturbing signal we have to make
infinitely many experiments and average over all possible responses. When the perturbing signal is random,
it is convenient to run Eq. (3) for infinitely many realizations of $\xi_{P}(t)$, and this, as we shall see hereby,
will force us to prepare the perturbing network $P$ as well as the perturbed network $S$.


In the general case of a dicothomous  renewal process,  $\xi(t)$, generated
with a waiting-times pdf  $\psi(t)$, the
probability density, that fixed a time
$t^{\prime}$, the first next event is observed at time $t>t^{\pe}$ is given \cite{jerry} by
\begin{equation}
 \label{renewgener}
 \psi(t,t^{\prime}) = \psi(t) +
 \sum_{n=1}^{\infty} \int_{0}^{t^{\prime }}R(t^{\prime \prime})
 \psi(t-t^{\prime \prime}) dt^{\prime \prime},
\end{equation}
with
\begin{equation}
\label{Pfun}
R(t)=\sum_{n=1}^{\infty} \psi_n(t),
\end{equation}
where $\psi_{n}(t)$ denotes the $n$-times convolution of $\psi(t)$.
$R(t)$ is therefore the probability density of having an event occurring exactly at time $ t $.
It  can be shown as well, see \cite{jerry},
that the auto-correlation function of the process is  related to $\psi(t,t^{\pe})$ by :
\BEQ
\label{autcorr}
\langle \xi(t) \xi(t^{\pe})\rangle=\int_t^{\infty}dx \psi(x,t^{\pe})=\Psi(t,t^{\prime})
\EEQ
and therefore coincides with the survival probability $\Psi(t,t^{\prime})$ for the first event,
i.e. the probability that, for fixed $t^{\pe}$, no event is  observed
until time $t>t^{\pe}$.
We  assume that the fluctuation $\xi_S(t)$ generated by the network $S$
is a dicothomous renewal process defined by the probability density
\begin{equation}
\label{waitingforS}
 \psi_S(t) =(\mu_S - 1) \frac{{T_S}^{\mu_S -1}}{(t + T_{S})^{\mu_s}}.
 \end{equation}
We therefore name respectively $\psi_S(t,t^{\pe})$, $R_S(t)$ and $\Psi_S(t,t^{\pe})$  the functions
 obtained
by replacing in Eqs. (\ref{renewgener}), (\ref{Pfun}) and (\ref{autcorr})  $\psi(t)$ with $\psi_S(t)$.

Let us consider now the Gibbs ensemble of networks $S+P$, and evaluate the
average $\langle\xi_S(t)\rangle_{SP}$. Note that the average is
over the separate statistics of the two networks $S$ and $P$

\begin{equation}\label{leadingarguments}
  \langle\xi_S(t)\rangle_{SP} =  \langle\langle\xi_S(t)\rangle_S\rangle_P.
\end{equation}
We select all the responses to the same perturbation,
characterized by  a given $\xi_{P}(t)$, we evaluate their average,
denoted by $\langle\xi_S(t)\rangle_S$, and finally we construct the
average over all possible perturbations denoted by
$\langle...\rangle_{P}$ so as to obtain the final result denoted by
$\langle...\rangle_{SP}$. In conclusion with this procedure we
obtain

\begin{equation}\label{notyetdefined}
\langle\sigma(t)\rangle = \langle\langle\xi_{S}(t)\rangle\rangle  = \epsilon \int_{0}^{t} dt^{\prime}
\chi(t,t^{\prime}) \langle\xi_{P}(t^{\prime})\rangle,
\end{equation}
where for notational convenience we drop the subscripts, but we
understand the averages in the sense described above.

If necessary, the signal $\xi_P(t)$ must share the same properties as $\xi_S(t)$ and for simplicity they are both assumed to be
 dichotomous signals with random renewal fluctuations
between the values $+1$ and $-1$. 
$\xi_{P}(t)$  is therefore a non-Poissonian dichotomic fluctuation
with the following waiting-time pdf:
\BEQ
\psi_P(t)=(\mu_P-1) \frac{T^{\mu_P-1}}{(t+T_P)^{\mu_P}}.
\EEQ
 It is therefore convenient to define the additional functions
$\psi_P(t,t^{\prime})$, $R_P(t)$ and $\Psi_P(t,t^{\pe})$
obtained, analogously as done for the   network $S$, by replacing in Eqs. (\ref{renewgener}), (\ref{Pfun}) and (\ref{autcorr})
the waiting time distribution $\psi_P(t)$.
The spectrum of this type of fluctuating signal, in the absence of perturbation, as calculated in Refs. \cite{mb,mirko}, is:
  \begin{equation} \label{mirkoisverybright}
S(f) \propto L^{\mu-2} f^{\mu-3},
       \end{equation} valid for $\mu<2$, remarkably, even though  a stationary auto-correlation function cannot be defined in this case. In the case $\mu > 2$, $S(f) = A/f^{3-\mu}$, with $A$  independent of $L$, the length of the sequence under study. 
At this point it should be clear to the reader that  to get the important results of this paper on the transmission of the statistical properties of $P$ to $S$, we must use Eq. (15).  This leads us to give a prescription to define $\langle \xi_{P}(t)\rangle$. For simplicity's sake, we shall assume that  the perturbing network  $P$ as well as the perturbed network $S$ are prepared at $t = 0$. Thus,  in Eq. (15) we shall  replace $ \langle\xi_{P}(t')\rangle$ with $\Psi_{P}(t')$.

\section{Phenomenological approach}

In this section we study the  response of a complex network producing non-poissonian renewal fluctuations to a  perturbing network generating similar fluctuations
within the phenomenological approach.
We analyze both the average response and the input-output correlation, i.e.
the correlation between the perturbing (input) fluctuating signal and the  signal produced by the ``system''  network (output).
In the phenomenologic approach,
the waiting times between the  events generating the dichotomic fluctuations remain unchanged by the perturbation. The external perturbation introduces a bias so that when an event occurs the probability that the dichotomic variable changes or keeps its value  are slightly different.
The function
 $\chi(t,s)$ in this approach, is given by \cite{ultimoallegro,primogerardo}:
                  \begin{equation}
          \label{FDTb}
        \chi(t,t^{\prime}) =
 \frac{d\Psi_S(t,t^{\prime})}{dt^{\prime}} =
R_S(t^{\prime}) \Psi_S(t-t^{\prime}).     \end{equation}
          The function $\Psi_{S}(t,t^{\prime})$ is the auto-correlation function of $\xi_S(t)$, namely, the survival probability of age $t^{\prime}$, and $R_S(t)$ for the case of discrete signals considered here, is the rate
 at which events are produced by the network $S$ prepared at $t = 0$, \emph{i.e.} the bits per second encoded in $\xi_S(t)$.  This rate is time independent only in the Poisson  case. In the non-Poisson case it depends on time, thereby making $\Psi_{S}(t,t^{\prime})$ non-stationary.
 The brand new survival probability $\Psi_S(t) = \Psi_S(t, t^{\prime}$=$0)$, is given by \cite{barbi,ultimoallegro,primogerardo}
       \begin{equation}
\Psi_S \left( t\right) = (1+t/T_S)^{1-\mu_S},  \label{nonsur}
\end{equation}
from which the corresponding waiting-times pdf   $\psi_S(t)=-d \Psi_S(t)/dt $ is  derived.
       In the range of parameters $1<\mu_S< 3$  considered here, it is known \cite{brucephysrep} that:
\begin{align}
\label{fadingaway}
      & R_S(t) \approx
-\frac{\sin{\pi \mu_S}}{T_S} (T_S/t)^{2- \mu_S}
 \;\;\; \;\mbox{ for}\; &1 < \mu_S < 2\\
\label{fadingaway2}
&R_S(t) \approx
\frac{1}{\tau_S}
\left[1 + \left(T_S/t \right)^{\mu_S -2}\right]  \;\;\;  \;\mbox{ for}\;\; & 2 < \mu_S <3,
\end{align}
with $\tau_S=T_S/(\mu_S-2)$ the mean value of $\psi_S(t)$.

 When $\mu_S < 2$ the experimental preparation of $S$ induces a sequence of events,  whose rate $R_S$ tends to vanish for $t\rightarrow \infty$, yielding a perennial out-of-equilibrium condition, and an explanation of the death of linear response \cite{barbi,sokolov1,sokolov2,heinsalu1,heinsalu2,weron1,sokolovall,sushin} as well. In  fact, the response to a harmonic perturbation of frequency $f$ is proportional to $1/(ft)^{2-\mu_S}$ \cite{barbi}.
        In the case $2< \mu_S <3$, on the contrary, the preparation-induced cascade of events, in the limit $t \rightarrow \infty$, becomes stationary and virtually identical to that of a Poisson process. The theoretical analysis of this paper is done in the asymptotic time regime. Thus, we refer to the case $2< \mu <3$  as \emph{stationary}, in  contrast to the \emph{non-stationary} case $\mu \leq 2$
of perennial transition.
 Similarly to the rate of events $R_S(t)$ the  spectral
 intensity per unit time tends to vanish for $\mu < 2$ as an effect of
 increasing $L$ (see Eq. (\ref{mirkoisverybright})).
The ideal $1/f$
 noise condition, corresponding to $\mu=2$,  generates instead a logarithmic decrease of the spectral intensity with time, and consequently a spectrum
virtually independent of $L$.

\subsection{Average response to perturbation.}

As previously mentioned,  the non-stationary LRT (NSLRT) of Eq. (\ref{LRT}) rests on the preparation of $S$ at time $t=0$. 
 We apply the same preparation condition to $P$, thereby generating the cascades $R_{S}(t)$ and $R_P(t)$ described by Eqs. (\ref{fadingaway}) and (\ref{fadingaway2}), with the appropriate indexing. Under this condition the  relaxation of $\langle \xi_P(t)\rangle$ becomes identical to the survival probability  $\Psi_P\left( t\right) $.
Assuming the condition of Eq. (\ref{FDTb})  we have the following expression for the average response:
\begin{align}\label{fenres}
\langle \sigma(t)\rangle=\eps \int_0^t R_S(t') \Psi_S(t-t')\Psi_P(t')dt' .
\end{align}
The preparation of both $S$ and $P$ makes the average over many realizations of the  response ${\sigma}(t)$ to a given stimulus $P$  vanish for $t\rightarrow \infty$.

 {\it Stationary case: $2<\mu_S<3$.}

In this regime a finite time scale for the fluctuation $\xi_S$ exists 
and $R_S(t)$ reaches the constant value $1/\tau_S$. 
 The inverse power law relaxation  of $\Psi_S(t)$ allows us to approximate Eq.(\ref{fenres}) by replacing $R_S(t')$  with its value for $t'\simeq t$, i.e.:
\BEQ \label{intermAa}
 \langle \sigma(t)\rangle \simeq \epsilon
 \int_{0}^{t}dt^{\pe}\Psi_S(t-t^{\pe})\Psi_P(t^{\pe})/\tau_S,
\EEQ
which becomes exact for $t \to \infty$.
The asymptotic behavior of Eq. (\ref{intermAa}) is easily obtained in the Laplace domain:
\BEQ
\langle\hat{\sigma}(s)\rangle\simeq \epsilon\frac{1}{\tau_S} \frac{1-\hat{\psi}_S(s)}{s} \frac{1-\hat{\psi}_P(s)}{s},
\EEQ
which can be studied in the limit of small $s$.
In fact, since \cite{klafter}:
\begin{align}
\hat{\psi}(s)&\simeq 1-\tau s  +\Gamma(\mu-2)s^{\mu-1}  \;\;\;\mu>2
\end{align}
and
\begin{align}
\hat{\psi}(s)&\simeq 1+\Gamma(\mu-2)s^{\mu-1}\;\;\;  \mu<2,
\end{align}
it follows that for $1<\mu_P<2$ and  $2<\mu_P<\mu_S $,  the time-asymptotic behavior is $\langle \sigma(t)\rangle\sim t^{1-\mu_P}$,   
which is proportional to $ \langle \xi_P(t)\rangle $ for large $t$, meaning that the system $S$  ``inherits'' the relaxation properties of the perturbation $P$.

For $2<\mu_S<\mu_P $, instead, 
the asymptotic dominant term is
 $\langle \sigma(t)\rangle\sim t^{1-\mu_S}$, which is proportional to  the ordinary unperturbed relaxation to equilibrium  $\langle \xi_S(t)\rangle$, when an initial bias  for $\xi_S(t)$ is introduced.
We see therefore that  for $\mu_P<\mu_S$, when  $\xi_P(t)$ is slower than $\xi_S(t)$, the perturbation imposes on the network  its own relaxation properties
thereby allowing one to ``manage'' the complexity of a network, by using an appropriate stimulus. 

{\it Non-stationary case:  $1<\mu_S \leq 2$}.

In this regime, the network  $S$ violates the finite-time scale condition
 necessary 
for stationary dynamics and in fact $R_S(t)\propto t^{\mu_S-2}$, see Eq. (\ref{fadingaway}).
With such replacement in Eq. (\ref{fenres}), a convolution form appears
which can easily studied via a Laplace transformation.
In the Laplace domain  (see Appendix A for details om coefficients):
\BEQ
\label{avglapl}
\langle\hat{\sigma}(s \simeq 0) \rangle\simeq
a_S \; s^{\mu_S-2} +a_P \; s^{\mu_P-2},
\EEQ
which implies that if $\mu_P>2$ or if $1<\mu_S<\mu_P$  then $\langle \sigma(t)\rangle\sim t^{1-\mu_S}$.
 When $1<\mu_P<\mu_S$, we have  
$\langle \sigma(t)\rangle\sim t^{1-\mu_P}$: Also in this case the perturbing network $P$  forces onto $S$ its own relaxation properties to equilibrium.

\subsection{Input-Output correlation function}
\begin{table}[h!]
\begin{tabular}{|l||l|l|}
\hline
$\mu_{S\downarrow}$  $\mu_{P\rightarrow}$
 & $\;\; \;\; \;1< \mu_P \leq 2$ & $\;\;\; \;\;\;\;\;\;\; \;\;2<\mu_P< 3$ \\
\hline

$1{\textstyle{<}} \mu_S {\textstyle{\leq}}2$
& $\Phi_{\infty} = \zeta(\mu_S,\mu_P)$$^*$\hspace{0.6 cm}I & $\Phi_{\infty}=0$ \hspace{1.8 cm}II\\
$2{\textstyle{<}}\mu_S{\textstyle{<}}3$
 &   $  \Phi_{\infty}=1$ \hspace{1.7 cm} III  & $
\Phi_{\infty}=\frac{\mu_S-2}{\mu_S+\mu_P-4}\;\;\; $ \hspace{0.4 cm}IV \\
\hline
\end{tabular}
\caption{Summary of the asymptotic  values of
the cross-correlation function $\Phi(t)$ in the phenomenological case. $\;^*$ See Eq. (\ref{ctb1}).}
\label{table1}
\end{table}
We study
the cross-correlation (or input-output correlation)  function
between the network $S$ and the stimulus $P$
: $C(t)\equiv \langle \langle \xi_S\left( t\right) \xi _P\left( t\right) \rangle \rangle $ 
which is also used as an indicator of aperiodic stochastic resonance \cite{collins}.
Multiplying both sides of Eq. (\ref{LRT}) by $\xi_P(t)$ and averaging over the fluctuations of the perturbation $P$
  we obtain:
\begin{equation}
\Phi(t)\equiv C(t)/\eps
= \int_0^t dt^{\prime }R_S(t')\Psi_S(t-t')\Psi _P\left( t,t^{\prime }\right) .  \label{cross2}
\end{equation}
Note that both  Eq. (\ref{fenres}) and Eq. (\ref{cross2}) depend on the survival probability of network P, but in the former such survival probability  depends on the  single time $t^{\prime}$ whereas in  the latter it depends on both $t^{\prime}$ and $t$.
We limit ourselves to report the  results for the asymptotic value
$\Phi_{\infty}$ of $\Phi(t)$.
When  $\xi_{S}(t)$ and $\xi_{P}(t)$ are not stationary, {\it i.e.}  when  $1<\mu_S \leq 2$ and
$1<\mu_P \leq 2$,  Eq. (\ref{cross2}), in the limit
 $t \to \infty$, gives:
\begin{flalign}\label{ctb1}
\ & \Phi_{\infty}=\zeta(\mu_S,\mu_P)\equiv
\Gamma(\mu_S+\mu_P-2) \times \\
\nn &   \frac{_3F_2\left[\{\mu_P-1,\mu_P-1,\mu_P+\mu_S-2\},\{\mu_P,\mu_P\},1\right]}{\Gamma(2-\mu_P) \Gamma(\mu_P)^2 \Gamma(\mu_S-1)},
\end{flalign}
where $_3F_2$ is the generalized hypergeometric function. For more details see Appendix A.
In the case $2<\mu_P<3$, $\Phi_\infty$ is simply zero.

 \begin{figure}[hbh]
\includegraphics[height=6.8 cm, width=8.8cm]{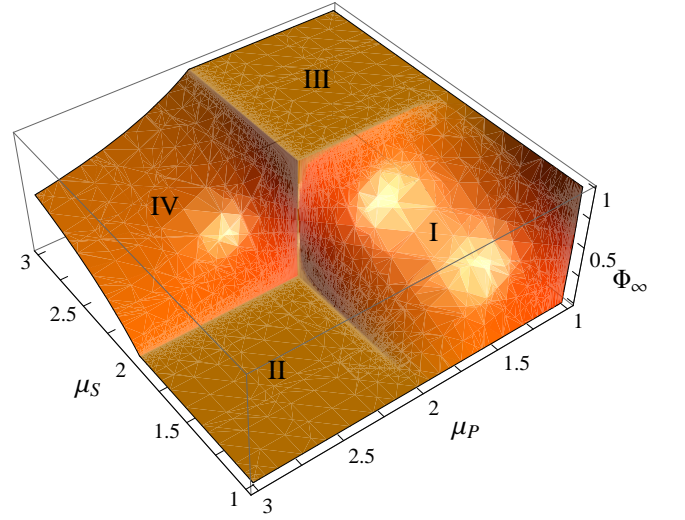}
%
\caption{The asymptotic limit of $\Phi(t)$ is displayed
for  $\mu_S,\mu_P \in ]1,3[$.
The vertex  $\mu_S$=$\mu_P$=$2$  marks the transition  to a condition of maximal input-output cross-correlation.}
\label{fig_1}
\end{figure}

In the case $2<\mu_S<3$,
inserting into  Eq. (\ref{cross2}) expression (\ref{fadingaway2}) for $R_S(t)$,
leads to:
\BEA
\label{correl_tc}
\nonumber
\Phi(t)    = \int^{t}_0 dt' \frac{{\Psi}_S(t - t')}{\tau_S} 
\Psi_P (t, t') =\tilde{\Psi}_S(0)\Psi_P (t, t)&&\\
-\tilde{\Psi}_S(t)	\Psi_P (t, 0)- \int^{t}_0 dt' \tilde{\Psi}_S(t - t') \frac{d \Psi_P (t, t') }{dt'},\;\;\;\;&& 
\EEA
where $\tilde{\Psi}_S(t)$ is given by  Eq. (\ref{nonsur}) after
replacing $\mu_S$ with $\mu_S-1$.
Eq. (\ref{correl_tc})  is exact for $t \gg \tau_S$ and for $1<\mu_P \leq 2$ it leads  to
$\Phi_{\infty} = 1$, since both the second and third term disappear for $t \to \infty$ and the first is trivially $1$ in the same limit. For  Eq. (\ref{correl_tc}) $2<\mu_P<3$ it yields:
\BEA
\label{statcorr}
&&\Phi_{\infty}=1-(\mu_P-2) T_P^{\mu_P-2} T_S^{\mu_S-2} \Delta T ^{4-\mu_S-\mu_P}\\
\nonumber&&\times  B[  \begin{array}{c}\Ss{\Delta T/ T_P}\\ \Ss{\Delta
T/T_S}\end{array} ,\mu_S+\mu_P-4,\begin{array}{c} \Ss{2-\mu_P}\\
\Ss{3-\mu_S}\end{array} ]\stackrel{\Delta T \to 0}{=}\frac{\mu_S-2}{\mu_P+\mu_S-4},
\EEA
where $\Delta T=|T_S-T_P|$, $B[x,a,b]$ is the incomplete Beta function
and the upper (lower) choice of the parameters   refers
to the case  $T_S> (<)T_P $. The final expression in Eq. (\ref{statcorr}) corresponds to the case $T_S=T_P$.
Results are summarized in
 Table \ref{table1}.
For illustrative purposes, we supplement Table I with  Fig. 1,  showing the  3D plot of
the cross-correlation function
 $\Phi_{\infty}$ in the same parameter range:
 Square II and square III correspond to the condition of minimal and maximal correlation, respectively. Intuitively it is so because of the difference of time scales between $S$ and $P$ in such regions. In III fluctuations $\xi_S(t)$ and $\xi_P(t)$ have a finite and an infinite time scale, respectively, thereby allowing
 $\xi_{S}(t)$ to adapt to the stimulus-induced bias so as to yield maximal correlation.  In  II the role of the time scales is inverted,
the bias induced by $P$  on the longer (diverging) time scale of the process $\xi_S(t)$ is asymptotically averaged out due to the many intervening switching events of $\xi_P(t)$, producing no correlation.
The vertex $\mu_S=\mu_P=2$, representing a $1/f$-noise network under the stimulus of a $1/f$-noise perturbation,  marks the abrupt transition from  vanishing (square II) to maximal correlation (III).


\section{ Dynamical approach}\label{transmission}
In this section  we extend the analysis of the previous section
to the dynamical approach.
Within such approach we derive both the average response and the
input-output correlation function.
\subsection{Average response to perturbation}
Starting from the  property

\begin{equation}\label{eq20}
\langle\sigma(t)\rangle = \langle\langle\xi_{S}(t)\rangle\rangle,
\end{equation}
obtained by averaging over the fluctuations of both networks and  using the dynamical condition
of Eq. (9), one obtains:
\begin{equation}
\chi(t,t')=-\frac{d}{dt}\Psi_S(t,t')=\psi_S(t,t')
\end{equation}
Eq. (\ref{LRT})
 then becomes:
\begin{equation}\label{meangeneral}
\langle \sigma(t) \rangle = \epsilon \int_{0}^{t} dt^{\prime}
\psi_S(t,t^{\prime}) \langle\xi_{P}(t^{\prime})\rangle.
\end{equation}

For simplicity we prepare the perturbation $\xi_{P}(t)$ at $t= 0$.
To observe the influence of $P$ on $S$, we select all the networks
of the Gibbs ensemble where $\xi_{P}(0) = 1$. In this case
$\langle\xi_{P}(t^{\prime})\rangle$ is given by \cite{jerry} the survival
probability $\Psi_P(t)$. Thus, Eq. (\ref{meangeneral}) yields

\begin{equation}\label{specificmeangeneral}
\langle \sigma(t) \rangle = \epsilon \int_{0}^{t} dt^{\prime}
\psi_S(t,t^{\prime}) \Psi_P(t^{\prime}),
\end{equation}
where
 $\Psi_P(t^{\prime})$ is the survival probability  for the process $\xi_P(t)$:
\begin{equation}\label{chidef}
\Psi_P(t) =\int_t^{\infty}dx\psi_P(x)= \left(\frac{T_P}{t+ T_P}\right)^{\mu_P-1}.
\end{equation}
This slow decay corresponds to the probability that no
perturbation event occurs up to time $t$. The network $S$ evolves
in time so as to reach a steady value that corresponds to a
constant perturbation abruptly applied at $t= 0$. However, during
this process a perturbation event occurs that has the effect of
suddenly changing the external field. Thus, $\sigma(t)$
 does not reach a steady value, but
after reaching a maximum value will decay. Under the specific
conditions discussed in this Section, the time asymptotic decay of
$\langle\sigma(t)\rangle$ in Eq. (\ref{specificmeangeneral}) has the same
power-law index as that of survival probability $\Psi_P(t)$. We
interpret this phenomenon as the transmission into $S$ of the
statistics of $P$.

We explore again the whole range of parameters  $1<\mu_S<3$ and $1<\mu_P<3$
respectively 
for the ``system''  network $S$
 and the perturbation $P$,  
depending on the values of the power-law indexes $\mu_S$,$\mu_P$ 
characterizing their  waiting-times pdfs.
The  value $\mu=2$ marks the transition from a finite to an infinite mean time, i.e. the transition to a  a non-ergodic, non-stationary condition. 
In fact, while for  $\mu>2$  a mean time exists, a finite time scale can be defined and a stationary condition is reached, for $\mu<2$ such condition is never achieved, not even in the infinite mean time.

{\it Stationary case: $2<\mu_S<3$. }

 In this regime the waiting-times pdf $\psi_S(t)$ has a finite mean  value $\tau_S$.
Therefore  a finite time scale $t_C$ exist such 
that for $t>t'>t_C$  the following approximation corresponding to reaching a 
stationary condition, is valid \cite{jerry}:
\begin{equation}\label{statpsi}
\psi_S(t,t^{\prime}) \simeq
\frac{1}{\tau_S}\int_t^{\infty}dx\psi_S(x-t^{\pe})=\frac{\Psi_S(t-t^{\pe})}{\tau_S},
\end{equation}
where $\tau_S=T_S/(\mu_S-2)$ is the mean value of
$\psi_S(t)$.
Eq. (\ref{statpsi}) is exact for $tc\to \infty$.
We can also use Eq. (\ref{LRT}) for the response
 in the case when the interaction is turned on at a later
time $t_C$ , that we  assume to be so large as to satisfy the approximation Eq. (\ref{statpsi}), using the following procedure. We introduce into Eq. (\ref{LRT}) an effective
perturbation  which  
is  turned on at time $t=0$ but   is
zero until $t_C$, i.e.  $\xi^{eff}_P(t)=\theta(t-t_C)\xi_P(t-t_C)$.
In this scheme we can  assume that the ``real" process  $\xi_P(t)$ is prepared at time
$t=t_C$  in a brand new condition. Time $t_C$ therefore corresponds to the age of the network $S$.
The average response $\langle\sigma(t)\rangle$ of network $S$  in this case reads:
\BEA\label{PIdecompose}
 \langle\sigma(t)\rangle&=&\eps
\int_0^{t}dt^{\pe}\psi_S(t,t^{\prime})\langle\theta(t^{\pe}-t_C)\xi_P(t^{\pe}-t_C)\rangle\\
\nonumber &=&\eps\int_{t_C}^{t}dt^{\pe}\psi_S(t,t^{\pe})\Psi_P(t^{\pe}-t_C),
\EEA
and the approximation  (\ref{statpsi}) can be used to replace $\psi_S(t,t')$.
With the substitution $\tau = t^{\prime} - t_C$ and after renaming
$ t -t_C$ as $t$  back again, 
 the average response of the network $S$ of age $t_C$  reads: 
\BEA \label{intermA}
\nonumber &&\frac{\langle\sigma(t)\rangle}{\eps}|_{t_C} =
\int_{0}^{t}
d \tau\frac{\Psi_S(t-\tau)}{\tau_S}\Psi_P(\tau)=\tilde{\Psi}_S(0)\Psi_P(t)\\\
&-&\tilde{\Psi}_S(t)\Psi_P(0)-\int_{0}^{t}dt^{\pe}\tilde{\Psi}_S(t-t^{\pe})\frac{d}{dt^{\pe}}\Psi_P(t^{\pe}),
\EEA
where $\tilde{\Psi}_S(t)$ is defined as:
\BEQ
\tilde{\Psi}_S(t)=\left(1+t/T_S\right)^{\mu_S-2}.
\EEQ
Eq. (\ref{intermA}) is exact for $t_C \to \infty$ and coincides 
with the expression Eq. (\ref{intermAa}) obtained in the phenomenological case in the same regime.
Therefore, in the limit $t>>T_S,T_P$, the same considerations apply, i.e.
if $1<\mu_P<2$ and  $2<\mu_S<\mu_P $,  the time-asymptotic behavior is $\langle \sigma(t)\rangle\sim t^{1-\mu_P}$,   
which is proportional to $ \langle \xi_P(t)\rangle $ for large $t$, therefore the network $S$ always ``inherits"
the relaxation properties of  the perturbation.
 If  $2<\mu_S<\mu_P $, 
 the asymptotic dominant term is 
 $\langle \sigma(t)\rangle\sim t^{1-\mu_S}$, which is proportional to  the ordinary unperturbed relaxation to equilibrium  $\langle \xi_S(t)\rangle$.


{\it Non-stationary case: $1<\mu_S<2$.}

Let us make the assumption that, although at time $t=0$ half of
the $S$ networks are in the state $\xi_S=+1$ and half in the state
$\xi_S=-1$, all of  them are at the beginning of their sojourn in
the corresponding states. This is an out of equilibrium condition,
corresponding to preparing the network at $t= 0$.
 The calculations are
detailed in Appendix B. Using the dynamic theory we obtain

\begin{equation}\label{pi_din}
\frac{\langle\sigma(t) \rangle}{\epsilon} \approx
\frac{k_1(\mu_S,\mu_P)}{t^{\mu_P -1}} +
\frac{k_2(\mu_S,\mu_P)}{t^{\mu_P +1-\mu_S}},
\end{equation}
where the first coefficient is given by
\begin{equation}
\label{k1}
k_1(\mu_S,\mu_P) = \frac{{T_P}^{\mu_P-1} \sin (\pi \mu_S)
\Gamma(2-\mu_S)\Gamma(1-\mu_P+\mu_S)}{\pi \Gamma(3-\mu_P)}
\end{equation}
and the second coefficient is determined to be
\begin{widetext}
\begin{equation}
\label{k2}
k_2(\mu_S,\mu_P) = \frac{\left[\sin (\pi \mu_S)
\Gamma(3-\mu_S)\Gamma(1-\mu_P+\mu_S)\Gamma(2\mu_S -3) -
(2-\mu_S)\Gamma(2 \mu_S -\mu_P-1)\right]}{{T_P}^{1-\mu_P} {T_S}^{
\mu_S-2}\pi (\mu_P-2) (2-\mu_S)\Gamma(3-\mu_S)\Gamma(\mu_S -\mu_P)\Gamma(2
\mu_S-3)}.
\end{equation}
\end{widetext}

In this range of parameters the dominant term is always the first term in 
Eq. (\ref{pi_din}) which,  if $\mu_P<\mu_S$, is also slower than the unperturbed relaxation to equilibrium of $\xi_S$.
In the latter range, therefore, the network $S$  
relaxes to equilibrium inheriting  the same properties of the perturbing network $P$.

The result in Eq. (41) is of special interest since it discriminates between the two approaches in the non-stationary regime. It is this difference that allowed to determine
 that  liquid crystals \cite{liquid}  follow the prediction of the dynamical approach.  In fact, the phenomenological
approach disregards the influence of the perturbation on the
occurrence time of the $S$ events \cite{sokolov1,sokolov2}, while the
dynamical theory does not, thereby affording a criterion for
information transport that we judge to be a more appropriate
representation of the communication among complex networks with
$\mu < 2$. However, the equivalence between the phenomenological
and the dynamic theories in the case when the network $S$ is
infinitely aged (i.e. for $\mu_S>2$), indicates that generalization of FDT given by Eq.
(\ref{dynamical}), namely, the dynamical theory, becomes active only
when the network $S$ is in a far from equilibrium condition and
begins drifting towards equilibrium. 
Although equilibrium is never
reached when $\mu<2$, the correlation function
$\Psi_S(t,t^{\prime})$ tends to recover the property
$\Psi_S(t,t^{\prime}) = \tilde{\Psi}_S(t-t^{\prime})$ that makes the
phenomenological theory formally equivalent to the dynamical
theory.  
%

\subsection{Input-Output correlation function}\label{correlation}

Herein we study the asymptotic limit of the input-output correlation function 
    \begin{equation}
    \Phi_{\infty} \equiv \lim_{t\rightarrow \infty} C(t)/\epsilon,
    \end{equation}
within the  dynamical approach.
The input-output correlation function is again defined by the average
over the fluctuations in both the $S$ and $P$ networks:
    \begin{equation}
    C(t) \equiv \langle\xi_{S}(t)\xi_{P}(t)\rangle.
    \end{equation}

The  asymptotic limit of $C(t)$  is independent of the way the
`'system'' network and the perturbation are prepared, so we can use the prescription
leading to Eq. (\ref{LRT}) obtained assuming that both network $S$ and
perturbation $P$ are prepared at time $t=0$.
We therefore use  the same arguments as those yielding Eq.
(\ref{cross2}) and,   adapting them to the dynamic theory  we
obtain
    \begin{equation}
    \label{fundamentalC}
  \Phi(t)= C(t)/\epsilon =  \int_{0}^{t} dt^{\prime} \psi_S(t,t^{\prime}) \Psi_P(t,t^{\prime}).
    \end{equation}

{\it Non stationary case I: $\mu_S<2,\,\,\mu_P<2$.}

 We use Eq. (\ref{fundamentalC})
with the general expressions for $\psi_S(t,t^{\prime})$ and
$\Psi_P(t,t^{\prime})$ as obtained through
Eqs. (\ref{renewgener}) and (\ref{autcorr}), respectively.
In this case, taking the limit $t \rightarrow \infty$
yields (see Appendix B  for details)
\begin{align}\label{ctb}
\nonumber &\zeta_D=\lim_{t\to\infty}\Phi(t)=-\frac{\sin\pi\mu_P}{\pi}
 \frac{\Gamma(\mu_P+\mu_S-1)}{(\mu_P-1)\Gamma(\mu_P+1)\Gamma(\mu_S-1)}\\
 &\times F\left[\{\mu_P-1,\mu_P-1,\mu_P+\mu_S-1\},\{\mu_P,\mu_P+1\},1\right]. 
\end{align}

{\it  Non-stationary case II: $\mu_S<2,\,\,\mu_P>2$.}
 
In this case
we can assume that, when the interaction is turned on, the perturbing network
$P$ has already reached a stationary condition,
 so that in  Eq.
(\ref{fundamentalC}) $\psi_S(t,t^{\prime})$ is  given
again by Eq. (\ref{renewgener}) but $\Psi_P(t, t^{\prime}) =
\tilde{\Psi}_P(t-t^{\prime})$, with
\begin{equation}
\label{Xistat}
\tilde{\Psi}_P(t)=(1+t/T_P)^{\mu_P-2},
\end{equation}
and this expression has to be  directly inserted into  Eq. (\ref{fundamentalC}).

The power-law index $\mu_P-2$ reflect the stationary condition
realized with the preparation  of the perturbation $P$ at a time $t_P=-\infty$.
In this case, we obtain
\begin{equation}\label{ctlimc_2}
  \lim_{t\to\infty}\Phi(t)=0.
\end{equation}
In the time asymptotic limit, the network $S$ turns out to be
independent of $P$ in spite of the fact that at $t=0$ we switch on
the $S-P$ interaction.

{\it Stationary case I: $\mu_S>2,\,\,\mu_P>2$.}

We again  use  Eq. (\ref{fundamentalC}) and assume that, when the interaction
is turned on, the perturbing network $ P$ has already reached
a stationary condition so   that  in (\ref{fundamentalC})
$\Psi_P(t,t^{\prime})=\tilde{\Psi}_P(t-t^{\prime})$ with $ \tilde{\Psi}_P(t)$ given by Eq. (\ref{Xistat}).
 For $\mu_S>2$ a finite mean time
$\tau_S$ of $\psi(t)$ exists, therefore 
a finite time scale   $t_{C}\propto \tau_S$ exists such that for $t>t_C$  Eq.(\ref{statpsi})  can be used again to approximate  $\psi_S(t,t')$. 
With such substitutions, the expression for the correlation function
becomes asymptotically equal to the correlation  obtained in the phenomenological approach in the same regime, i.e. Eq. (30), leading to the final result  Eq. (\ref{statcorr}).  

{\it Stationary case II: $\mu_S>2,\,\,\mu_P<2$.}

In this case again a finite time scale  $t_C$ can be found such that the
approximation in Eq. (\ref{statpsi}) is valid.
 Thus,  again
Eq. (\ref{fundamentalC}) can be rewritten
\BEA\label{chi_b}
\label{decomposed2}
\Phi(t)&\simeq&\int_0^{t_C}dt^{\prime}\psi_S(t,t^{\prime})\Psi_P(t,t^{\prime})
\\
\nonumber &&+\int_{t_C}^{t}dt^{\prime}[\frac{d}{dt^{\prime}}\tilde{\Psi}_S(t-t^{\prime})] \Psi_P(t,t^{\prime}).
\EEA

This case is therefore equivalent to what obtained for the phenomenological case in the same range of parameters.
Again, in the asymptotic limit $t \to \infty$, the first term in
Eq. (\ref{decomposed2}) vanishes and  after integrating the second term  by parts, one obtains:
\BEA
\label{3terms}
\nonumber \Phi(t)&\simeq&  \tilde{\Psi}_S(0) \Psi_P(t,t)
-\tilde{\Psi}_S(t-t_C)\Psi_P(t,t_C)\\
 &-&\int_{t_C}^{t}dt^{\pe}\tilde{\Psi}_S(t-t^{\pe})\frac{d \Psi_P(t,t^{\pe})}{dt'}
\EEA
\begin{table}[h!]
\begin{tabular}{|l||l|l|}
\hline
$\mu_{S\downarrow}$  $\mu_{P\rightarrow}$
 & $\;\; \;\; \;1< \mu_P \leq 2$ & $\;\;\; \;\;\;\;\;\;\; \;\;2<\mu_P< 3$ \\
\hline

$1{\textstyle{<}} \mu_S {\textstyle{\leq}}2$
& $\Phi_{\infty} = \zeta_D(\mu_S,\mu_P)$$^*$\hspace{0.6 cm}I & $\Phi_{\infty}=0$ \hspace{1.8 cm}II\\
$2{\textstyle{<}}\mu_S{\textstyle{<}}3$
 &   $  \Phi_{\infty}=1$ \hspace{1.7 cm} III  & $
\Phi_{\infty}=\frac{\mu_S-2}{\mu_S+\mu_P-4}\;\;\; $ \hspace{0.4 cm}IV \\
\hline
\end{tabular}
\caption{Summary of the asymptotic  values of
the cross-correlation function $\Phi(t)$ in the dynamical case. $\;^*$ See Eq. (\ref{ctb}).}
\label{table2}
\end{table}

\begin{figure}[bhb]
\includegraphics[height=6.8 cm, width=8.8cm]{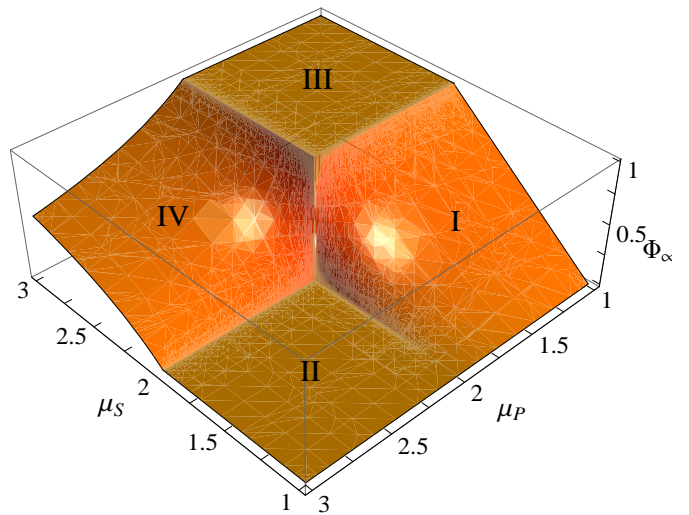}
\caption{The asymptotic limit of $\Phi(t)$ is displayed
for the different regimes of parameters for $1<\mu_S<3$ and $1<\mu_P<3$, in the dynamical approach.
}
\label{fig_00}
\end{figure}

In the asymptotic limit the second term on the right-hand side  trivially vanishes
and also the third term  side can be shown to vanish (see
Appendix B). The only remaining contribution in Eq. (\ref{3terms}) is given by the first term, which is exactly one.
It follows:
\begin{equation}\label{a7}
\Phi_{\infty} = \lim_{t\to\infty}C(t)/\epsilon=1.
\end{equation}
Let us make here a few remarks on these results.
Of course in the absence of coupling, $\epsilon = 0$,  which means that $C(t)$ always vanishes. We switch the
interaction on at $t= 0$. Thus, we always have the zero
cross-correlation initial condition
%
 $  C(0) = 0.$
As an effect of switching on the interaction at $t=0$ we realize
the condition
%
 $  C(t) \neq 0$,
for $t> 0$. However, we find that there exist special
conditions (region II in Fig. 2) for which the cross-correlation again goes to zero,
asymptotically
   \begin{equation}
  \Phi_{\infty} =\lim_{t\to \infty}C(t)/\epsilon= 0.
   \end{equation}
This indicates that only in such conditions, the network $S$, after a transient, recovers
the condition of statistical independence of the perturbation
$\xi_{P}(t)$. The fluctuations  $\xi_S$ with $\mu_S < 2$ turn out to be
statistically independent of $\xi_P(t)$ only when $\mu_P>2$.  The environmental perturbation with $\mu_P<2$, on the other
hand, is characterized by the remarkable property of forcing the
network $S$ to respond, regardless of the value of $\mu_S$.

Table II and the  plot in Fig. 2 summarize such results and 
 show the same qualitative properties observed
for the correlation function in the phenomenological case, with the correlation functions for the two approaches being identical in squares $I,II,IV$.
The condition of ideal $1/f$-noise, i.e. $\mu_S=\mu_P=2$   marks 
the transition from a condition of zero to maximal correlation.

\section{Finite Response and $1/f$-resonance}
Herein
 we proceed to demonstrate   that the intensity of the response
$\sigma(t)$ to a single realization of the stimulus
 does not decay, if $\Phi_{\infty} \neq 0$. This demonstration, therefore,
is valid for both the dynamical and the phenomenological approach, leading
to a general result.
Let us define with $p_t(\xi_S^i)$ the probability that at time $t$
the variable $\xi_S$  takes the value $i=\pm 1$ and with $p_t(\xi_S^i|\xi_P^j)$
 the conditional probability for the occurrence, at time $t$, of a value $\xi_S\textstyle{=}i\textstyle{=}\pm 1$, given the occurrence of a  value  $\xi_P\textstyle{=}j\textstyle{=}\pm 1$.
By definition, the non-vanishing $\Phi_{\infty}$ yields:
\BEQ
\label{formalcorr}
C(t) \equiv \sum_{i,j=\pm 1} i  j\; p_t(\xi_S^i|\xi_P^j) \; p_t(\xi_P^j)
\to \eps \Phi_{\infty}.
\EEQ
We note that for $t \to \infty$,  on a time scale such that 
$\langle \xi_P(t)\rangle$, which decays as $\langle \xi_P(0)\rangle t^{1-\mu_P} $, 
is a second-order quantity, $O(\eps^2)$, we have that:
\begin{align}
\label{pp}
 p_t(\xi_P^j)= 1/2+O(\eps^2)
\end{align}
and
\begin{align}
\label{phiINF}
 \Phi(t)=\Phi_{\infty}+O(\eps^2).
\end{align}
  Thus, due to Eq. (\ref{formalcorr}) and to the symmetry of the considered dichotomous processes:
\begin{flalign}
\label{conditional}
p_t(\xi_S^i|\xi_P^j)
\to \frac{1}{2} + i \, j \eps \frac{\Phi_{\infty}}{2}.
\end{flalign}
In the same long-time scale, Eq. (\ref{conditional}) yields:
\BEQ
\label{finalarg0}
\langle \sigma(t)\rangle_{\pm}
\equiv \sum_{i}
p_t(\xi_S^i|\xi_P^{\pm 1})
\; i \simeq  \pm \eps
\Phi_{\infty},
\EEQ
where the subscript $\pm$ indicates the value of $\xi_P$ at time $t$.
Summing Eq. (\ref{finalarg0}) over the two values of $\xi_P$, gives a total average null response,
 as expected.
<But if the magnitude $|\sigma(t)|$ of the response to a single instance of the input $\xi_P(t)$ is considered instead,  its total average is:
\begin{align}
\label{finalarg}
\langle |\sigma(t)|\rangle=\frac{1}{2} \sum_{\pm}\langle|\sigma(t)|\rangle_{\pm}
\gtrsim \frac{1}{2} \sum_{\pm}|\langle\sigma(t)\rangle_{\pm}|
\simeq\eps \Phi_{\infty},
\end{align}
where an equality holds if terms of order $O(\eps^2)$ are neglected.
Thus when $\Phi_{\infty} > 0$, the response $\sigma(t)$ to a single instance of the input $\xi_{P}(t)$ \emph{does not die out and remains proportional to the stimulus intensity}, no matter how large $t$ becomes. Square III in both Figs. 1 and 2, is the plateau region of maximal cross-correlation and response.
 Claims regarding the \emph{death of linear response}, as in \cite{sokolov1,sokolov2,heinsalu1,heinsalu2,weron1,sokolovall, sushin, magiaro,kang,australians},  are therefore appropriate only in relation to the vanishing correlation of square II.
The total average response
$\langle\sigma(t) \rangle$  always tends to vanish
   for  $t \rightarrow \infty$  for reasons that do not imply a lack of response except in the case of square II. 

The reason for the striking difference between the response to a harmonic perturbation and the response to a non-ergodic stimulus is intimately related to the emergence of $1/f$ noise and to its spectrum described by Eq. (\ref{mirkoisverybright}) which
assigns the  weight $S(f)/L=1/(fL)^{3-\mu_P}$ to the spectral component of frequency $f$ of a non-ergodic stimulus.
As a consequence, the stimulus generates, in time, lower and lower  frequencies $f$, so as to keep  $1/(fL)^{2-\mu_S}$ ({\it i.e.} the response intensity to frequency $f$ \cite{barbi}) finite, thereby yielding Eq. (\ref{finalarg}). The death of linear response   
\cite{sokolov1,sokolov2,heinsalu1,heinsalu2,weron1,sokolovall, sushin, magiaro,kang,australians}
 is caused by the fact that
stimuli with fixed frequencies cannot cope with the decreasing frequency of the cascade of events of Eq. (\ref{fadingaway}).

We have afforded a compelling proof that the intensity of the single realizations of $\sigma(t)$, with $\mu_S < 2$, does not decay if the perturbation $\xi_{P}(t)$ falls in the same complexity basin ($\mu_P < 2$).
This is the phenomenon of 
{\it complexity management}
which allows to define the right stimulus to obtain a response from a 
network with non-ergodic properties.
Now we argue that $1/f$ stimuli generate the maximum information transport by
looking at
the mutual information
\BEQ
\label{mutual}
 I(t)=\sum_{i,j}p_t(\xi_P^j)p_t(\xi_S^i|\xi_P^j)\log[p_t(\xi_S^i|\xi_P^j)/p_t(\xi_S^i)].\EEQ
Using  Eqs. (\ref{pp}), (\ref{phiINF}) and (\ref{conditional})  it follows that:
\BEQ 
I(t \to \infty)\simeq\eps^2 \Phi^2_{\infty}
\EEQ
 and the information transmission rate is obtained by multiplying  $I(t)$
by the input rate \cite{shannon}, given by $R_P(t)$.
If $\mu_{P} > 2$, Fig. 1 shows that $\Phi_{\infty} < 1$.
Although square  III in  Figs. 1 and 2,  indicates  that  all stimuli with $\mu_{P} \leq  2$ induce maximal correlation,
 $\mu_P<2$ corresponds to
a stimulus   with decaying events rate (input bits/sec) $R_P(t)$.
 So even if a response is produced in this regime,  the rate of  information vanishes in time.
 Only at the crucial condition $\mu_P=2$, of ideal $1/f$-noise, does  this algebraic decay becomes logarithmic, and, consequently, a steady  and maximal information transmission rate is achieved. This is the phenomenon that we 
call $1/f$-resonance.
The above consideration are valid for both the dynamical and phenomenological approaches,
therefore we consider the  condition of maximal information transmission achieved in the ideal $1/f$-noise condition, a fundamental property of $1/f$-noise renewal processes.


\section{Concluding Remarks}\label{concludingremarks}
The growing interest for the dynamics of complex networks
is shifting the attention of the researchers from the synchronization
of two stochastic units \cite{pecora} to the synchronization of a large number of units \cite{tutorial}, an interesting phenomenon that is closely related to the very popular model 
of Kuramoto \cite{kuramoto}. The single units of the processes of chaos synchronization are chaotic and do surprisingly synchronize while maintaining the erratic 
dynamics that they have in isolation. Although the single units of the Kuramoto model
are regular, it is becoming increasingly evident that the emergence of a global synchronization is a condition independent of whether the single units are regular or stochastic. The single units of the work of Refs. \cite{bianco,gosia} are Poisson processes and if one of them drove the other, they would obey the principle of aperiodic stochastic resonance \cite{mirkofronzoni}.  If the two units are bi-directionally coupled they are expected to undergo a condition of perfect synchronization if the coupling is sufficiently intense. When the number of interacting units is very large a phase transition occurs from the non-cooperative to the cooperative behavior \cite{bianco,gosia}.  It is important to stress that at criticality
no permanent consensus is reached, and the mean value of the global field vanishes.
Yet, this condition is strikingly different from the non-cooperative condition. The whole network remains in the $``yes"$ ($``no"$) state for an extended time before making a transition to the $``no"$ ($``yes"$) state. 

It is surprising that the phase-transition literature seems to have overlooked, with only a few exceptions \cite{bianco,gosia,greeks}, that the transitions from the $``yes"$ ($``no"$) to the $``no"$ ($``yes"$) state  occurring at criticality are the ``crucial" events defined in Section \ref{Newintroduction}. In other words, the time interval   between two consecutive transitions is derived from a pdf  that has the asymptotic time structure of Eq. (\ref{specialform}) with a power index $\mu$ fitting the inequality condition of Eq. (\ref{crucial}). Some authors  \cite{bianco,gosia} argue that $\mu = 1.5$ and others,
\cite{notredame}, releasing the condition  that all the units share the same Poisson rate, generate a global condition with crucial events characterized by $\mu < 2$, but significantly departing from the value $\mu = 1.5$. Note  that the theoretical arguments of Ref. \cite{gosia2}, yield the misleading impression that the crucial value of $\mu$ is a consequence of ordinary statistical physics.   

According to some authors \cite{innerouter,outer} the Kuramoto phenomenon can be defined as \emph{inner synchronization}. A network of cooperating units located on the nodes of a complex network may reach inner synchronization with different values of the control parameter, depending on the network topology \cite{perc}. This is a subject of increasing interest
with attractive applications to the dynamics of the human brain \cite{innerouter}. If we adopt this perspective, we can address the problem of information transmission from one to another complex network as a process of \emph{outer synchronization}. This is an interesting issue, but the conditions to fulfill to realize outer synchronization are not yet clear \cite{innerouter}. 

An important result of this article is the discovery of a promising road to settle the problem of information transmission from one to another complex network. In fact, if the inner synchronization corresponds to a criticality condition and criticality generates crucial events with a power-law index $\mu$ fulfilling the inequality of Eq. (\ref{crucial}), then a complex network at criticality is a generator of $1/f$ noise, with $S(f) \propto 1/f^{3-\mu}$. 
Thus the problem of information transmission from one to another complex network becomes equivalent to the phenomenon of $1/f$-resonance illustrated in this article. This is essentially the main result of  earlier work  \cite{pinknoise}. The main conclusions of Ref. \cite{pinknoise} are illustrated by Fig. 1, which is obtained using the phenomenological LRT.

What are the limits of this earlier result? The experiments \cite{cascade,liquid} yield support to the dynamical rather than  
the phenomenological LRT, thereby generating  doubt that the results 
of Ref. [27], although very attractive, may not completely reflect reality. 
It is important to stress that  phenomenological LRT is a natural consequence of adopting the asymptotic time perspective replacing the
waiting-times pdf  $\psi(\tau)$ of Eq. (\ref{specialform})
with $\psi(\tau) \propto 1/\tau^{\mu}$. This way of proceeding, although
generating the elegant mathematics of fractional derivatives, has as an ultimate effect the misleading discovery of the death of linear response 
\cite{sokolov1,sokolov2,heinsalu1,heinsalu2,weron1,sokolovall, sushin}.
 We do not adopt the asymptotic time perspective but the special form of Eq. (\ref{specialform}).
 This is not a unique way of connecting the long-time to the short-time regime. However, whatever form we adopt we are convinced that there will be a parameter playing the same microscopc role of the parameter $T$ of Eq. (\ref{specialform}).  It is reasonable to assume that an external perturbation may perturb either $T$ or $\mu$, or both of them. However, the perturbation of $\mu$ is incompatible with the assumption of a weak stimulus. In fact, $\mu$ is a consequence of the cooperation among the units of the network, and a perturbation may affect $\mu$ only if its strength is large enough to influence the interaction among the units of the network. Thus, an external weak perturbation can only have an effect on $T$, thereby making the dynamical LRT become the proper way to study the response of a complex network to a weak external stimulus, in accordance with the experimental results \cite{cascade,liquid}.
 
 For these reasons, we can conclude that Fig. 2 is the original, and important, result of this paper.
 We hope that it may open the road to the dynamical solution of the problem of information transmission \cite{alexbialek} from one to another complex network, a research topic that is still in its infancy.

\emph{acknowledgments} PG acknowledges financial support from ARO
and Welch through grants W911NF-05-1-0205 and B-1577,
respectively.

\renewcommand{\theequation}{A-\arabic{equation}}
\setcounter{equation}{0}  
\section*{APPENDIX A}
In this Appendix we record more details about the derivation
of both the average response $\langle \sigma (t)\rangle$ and the  input-output correlation function $\Phi(t)$, in the case of the phenomenological approach.

When we adopt the phenomenological theory, we obtain for the average response to 
external perturbation, in the non-stationary case (cfr. Eq. (27)) the following asymptotic expression:

\begin{equation}\label{pi_phenom}
\frac{\langle\sigma(t) \rangle}{\epsilon} =
\frac{k_1(\mu_S,\mu_P)}{t^{\mu_P -1}} +
\frac{k_2(\mu_S,\mu_P)}{t^{\mu_S -1}},
\end{equation}
where the coefficient of the first term is

\begin{equation}
k_1(\mu_S,\mu_P) = \frac{\Gamma(\mu_P -
\mu_S)}{\Gamma(1-\mu_P)\Gamma(\mu_S)} - \frac{\Gamma(1 -\mu_P +
\mu_S)}{\Gamma(2-\mu_P)\Gamma(\mu_S)}
\end{equation}
and the coefficient of the second term

\begin{equation}
k_2(\mu_S,\mu_P) = \frac{\Gamma(\mu_P -
\mu_S)}{\Gamma(1-\mu_S)\Gamma(\mu_P)} .
\end{equation}
Note the logarithmic corrections corresponding  to $\mu_S =
\mu_P$, with

\begin{equation}
\frac{\langle\sigma(t)\rangle}{\epsilon} \approx \frac{\sin (\pi
\mu_S)}{\pi} \frac{\log t}{t^{\mu_S -1}} +
\frac{A(\mu_S)}{t^{\mu_S -1}},
\end{equation}
and

\begin{equation}\label{amus}
A(\mu_S) =\frac{\sin (\pi
\mu_S)}{\pi}\left[\frac{1}{\mu_S-1}-2\gamma-2
\psi_L(\mu_S)\right]-\cos (\pi \mu_S)
\end{equation}
where $\gamma$ is the Euler's constant and $\psi_L(z)$ the
logarithm derivative of the $\Gamma$ function. These predictions
are qualitatively equivalent to the dynamical theory predictions
with the assumption that the network $S$ has been prepared in the
very distant past.

The  general expression of the correlation function
in the phenomenological approach is given by Eq. (28).

{\it Non-stationary case I:  $\mu_S<2, \mu_P<2$.}

In this range the asymptotic approximation for the function
$R(t)$ defined in Eq. (\ref{Pfun}) for both $S$ and $P$ is
\BEQ
\label{appr}
R(t)\simeq \frac{\sin \pi \mu}{\pi}t^{\mu-2}
\EEQ
The power-law properties of $\Psi_S(t)$, in the long time limit,
shift the dominant contribution to the integral of Eq. (28)
 to the range  $t' \sim t$, this allows to adopt the approximation (\ref{appR}) inside the integral.
It follows
\begin{align}
\Phi(t)\simeq &\frac{\sin \pi\mu_S}{\pi}\frac{\sin \pi \mu_P}{\pi}\times\\
\nn &\times \int_0^t \tau^{\mu_S-2} \Psi_S(t-\tau)d\tau \int_0^\tau x^{\mu_P-2}\Psi_P(t-x).  
\end{align}
Using a generalized Newton binomial expansion of the power-law form of
the functions $\Psi_P(t)$ we obtain the following expression:
\begin{align}
\nn \Phi(t)\simeq&\frac{\sin \pi\mu_S}{\pi}\frac{\sin \pi \mu_P}{\pi} \sum_{n=0}^{\infty}{1-\mu_P \choose n}\frac{(-1)^n}{(t+T_P)^{n+\mu_P-1}}\\
 &\int_0^t t'^{n+\mu_P+\mu_S-3}\Psi_S(t-t')dt'
\end{align}
while the convolution with $\Psi_S(t)$ in the integral, leads to
\begin{align}
\nn \Phi(t)\simeq&\frac{\sin \pi\mu_S}{\pi}\frac{\sin \pi \mu_P}{\pi} \sum_{n=0}^{\infty}{1-\mu_P \choose n}\frac{(-1)^n}{(t+T_P)^{n+\mu_P-1}}\\
 &\frac{t^{n+\mu_P-1}}{n+\mu_P-1}\frac{\Gamma(n+\mu_P+\mu_S-2) \Gamma(2-\mu_S)}{\Gamma(n+\mu_P)}
\end{align}
where we have used the fact that for $t$ large the main contribution  to the integral comes from  the range of values $ x\lesssim \tau  \lesssim t$
and therefore the asymptotic approximations for $R_S(t)$ an $R_P(t)$ are justified.
The limit for $t\to \infty$ cancel the time dependence and leaves a sum which leads to the expression of Eq.(\ref{ctb}) in the text.

{\it Non-stationary case II:  $\mu_S<2, \mu_P>2$.}

In this case the function $R_P(t)$ tends asymptotically to a constant value
and therefore it easy to see, following analogous procedure to the previous case, that the correlation tends to zero.

{\it Stationary case I: $\mu_S>2, \mu_P<2$.}

As mentioned in the text, from Eq. (30) it is enough to show that the third
term on the right-hand side is zero. Such terms read as:
\BEQ
\frac{\sin \pi \mu_P}{\pi}\int_{0}^t \tilde{\Psi}_S(t-t')dt't'^{\mu_P-2}\Psi_P(t-t')
\EEQ
Using  a generalized binomial expansion for both $\tilde{\Psi}_S(t)$ and $\Psi_P(t)$ we get:
\begin{align}
\nn& \Phi(t)\simeq\frac{\sin \pi \mu_P}{\pi}\sum_{n,m}{1-\mu_P \choose n}{1-\mu_S \choose m}\frac{(-1)^n}{(t+T_P)^{n+\mu_P-1}}\\
& \times \frac{(-1)^m}{(t+T_S)^{m+\mu_S-2}} \int_0^t {t'}^{n+m+\mu_P-2} dt'
\end{align}
Carrying out the integration shows that the leading term in $t$ vanishes
for $t \to \infty$
Such demonstration is valid also for the dynamical case in the same
range of parameters (see Eqs. (\ref{3terms}) and (\ref{a7})).

\section*{APPENDIX B}
\renewcommand{\theequation}{B-\arabic{equation}}
\setcounter{equation}{0}  

This Appendix is devoted to  detailed calculations involved the
evaluation of the  asymptotic limits of the function $\Phi(t)$ and $\langle \sigma(t)\rangle$ in the dynamic approach.

 Herein only  the parameter range $\mu_S<2, \mu_P<2$ is considered, the other ranges being analyzed in detail in the main text and in Appendix A.
 
$\Phi(t)$ is given by the following integral:  
\begin{equation}\label{cnstpn1_bb}
  \Phi(t)=\int\limits_{0}^{t}\psi_S(t,t')\Psi_P(t,t')dt'
\end{equation}
where $\psi_S(t,t')$ is defined in  Eq. (\ref{renewgener})  and $\Psi_P(t,t')$
 in Eq. (12), each with the appropriate labeling. $\Phi(t)$ can be decomposed as sum of  four contributions:

\begin{equation}\label{ct}
\Phi(t)=\sum_{i=1}^{4}\Phi_i(t)
\end{equation}
where
\begin{equation}\label{c1}
\Phi_1(t)= t\psi_S(t)\Psi_P(t)
\end{equation}
\begin{align}\label{c2}
\Phi_2(t)&= \psi_S(t)\int\limits_{0}^{t}\int\limits_{0}^{t'}
  R_P(t'')\Psi_P(t-t'')dt''dt'\\
\nn &=\psi_S(t)\int\limits_{0}^{t}(t-t'')
  R_P(t'')\Psi_P(t-t'')dt''
\end{align}
\begin{align}\label{c3}
\Phi_3(t)&= \Psi_P(t)\int\limits_{0}^{t}\int\limits_{0}^{t'}
  R_S(t'')\psi_S(t-t'')dt''dt'\\
\nn &=\Psi_P(t)\int\limits_{0}^{t}(t-t'')
  R_S(t'')\psi_S(t-t'')dt''
\end{align}
\begin{equation}\label{c4}
\Phi_4(t)=
\int\limits_{0}^{t}\int\limits_{0}^{t'}\int\limits_{0}^{t'}
  R_S(t'')\psi_S(t-t'')R_P(\tau)\Psi_P(t-\tau)dt'dt''d\tau
\end{equation}
Here $\psi_S(t)$ and $ \Psi_P(t)$ are defined in
Eqs.(\ref{waitingforS}) and (\ref{chidef}).
Using methods of Ref.
\cite{wbg}, Eqs. (\ref{c2}), (\ref{c3}) can be evaluated in the
asymptotic limit to yield:
\begin{equation}\label{c1b}
\nn \Phi_1(t)\approx \frac{c_1}{t^{\mu_S+\mu_P-2}},\,\,\,\,\Phi_2(t)\approx
\frac{c_2}{t^{\mu_S-1}},\,\,\,\,\Phi_3(t)\approx
\frac{c_3}{t^{\mu_P-1}}.
\end{equation}
Since in the limit $t\to\infty$ the three contribute vanish, it is
not important the evaluation of the constants $c_1$, $c_2$, $c_3$.
To evaluate $\Phi_4(t)$ we first need an analytical expression for
$R_S(t)$ and $R_P(t)$. It can be shown that \cite{wbg}
\begin{equation}\label{p}
R_S(t)\approx -\frac{ \sin(\pi \mu_S )}{\pi
T_S^{\mu_S-1}}\frac{1}{t^{2-\mu_S}}.
\end{equation}
with the same expression, with the respective parameters, being valid
for $R_P(t)$.
Then we have

\begin{equation}\label{psitt}
\nn \int\limits_{0}^{t'}
  R_S(\tau)\psi_S(t-\tau)d\tau\approx -\frac{(t+ T_S-t')^{1-\mu_S } t'^{\mu_S-1 } \sin(\pi \mu_S )}{\pi (t+T_S)}.
\end{equation}
On the other hand

\begin{align}\label{Psitt}
\int\limits_{0}^{t'}
 &R_P(\tau)\Psi_P(t-\tau)d\tau =\frac{T_P^{\mu_P-1}}{(t+T_P)^{\mu_P-1}}\times\\
\nn & \times \sum\limits_{n=0}^{\infty}{1-\mu_P
\choose n}\frac{(-)^{n}}{(t+T_P)^{n}}
 \int\limits_{0}^{t'}
 R_P(\tau)\tau^{n}d\tau.
\end{align}
Using for $R_P(t)$ the same approximation of $R_S(t)$ Eq. (B-2), we have

\begin{align}\label{Psitt2}
&\int\limits_{0}^{t'}
 R_P(\tau)\Psi_P(t-\tau)d\tau \approx-
 \frac{\sin\pi\mu_P}{\pi(t+T_P)^{\mu_P-1}} \times\\
 \nn &\sum\limits_{n=0}^{\infty}{1-\mu_P
\choose
n}\frac{(-)^{n}}{(t+T_P)^{n}}\frac{t'^{n+\mu_P-1}}{n+\mu_P-1}.
\end{align}
The function $\Phi_4(t)$ is
\begin{align}
&\Phi_4(t)\approx
 \nn \frac{\sin\pi\mu_P\sin\pi\mu_S}{\pi^{2}(t+T_P)^{\mu_P-1}}
 \frac{t^{\mu_S+\mu_P-1}}{(t+T_S)^{\mu_S}}
 \sum\limits_{n=0}^{\infty}\left(-\frac{t}{t+T_P}\right)^{n}\times\\
\nn &  {1-\mu_P
\choose n}
\frac{F\left(n+\mu_P+\mu_S-1,\mu_S-1,n+\mu_P+\mu_S,
\frac{t}{t+T_S}\right)}{(n+\mu_P-1)(n+\mu_P+\mu_S-1)}
\end{align}
where $F(a,b,c,z)$ is the hypergeometric function. Finally

\begin{align}
&\Phi_4(t)\approx
 \nn \frac{\sin\pi\mu_P \Gamma(\mu_P+\mu_S-1)(t+T_P)^{1-\mu_P}t^{\mu_S}}{(1-\mu_P)\Gamma(\mu_P+1)\Gamma(\mu_S-1)(t+T_S)^{\mu_S}t^{1-\mu_P}}\times\\
\nn &F\left[\{\mu_P-1,\mu_P-1,\mu_P+\mu_S-1\},\{\mu_P,\mu_P+1\},
\frac{t}{t+T_S}\right]
\end{align}
where $_{p}F_{q}(\{a\},\{b\},z)$ is the generalized hypergeometric
function. In the limit for $t\to\infty $

\begin{align}\label{ctb1}
\Phi_{\infty}=&-
 \frac{\sin\pi\mu_P \Gamma(\mu_P+\mu_S-1)}{ \pi (\mu_P-1)\Gamma(\mu_P+1)\Gamma(\mu_S-1)}\times\\
\nn & F\left[\{\mu_P-1,\mu_P-1,\mu_P+\mu_S-1\},\{\mu_P,\mu_P+1\},1\right],
\end{align}

As far as the evaluation of the function $\langle \sigma(t)\rangle$ is concerned, in the dynamic approach it is
given by Eq. (\ref{specificmeangeneral}):
\begin{equation}\label{tp1}
\langle \sigma(t)\rangle=\epsilon\int_{0}^{t}\psi_S(t,t')\Psi_{P}(t')dt',
\end{equation}
 which is to be evaluated in the parameter range $\mu_S<2, \mu_P<2$.
Let $P_S(t)=R_S(t)+\delta(t)$,  Eq. (\ref{tp1}) reads:
\BEQ
\label{intdin}
\langle \sigma(t)\rangle=\epsilon\int_0^t dt' \Psi_P(t') \int_0^{t'} dx P_S(x) \psi_S(t-x),
\EEQ
which, after a transformation on the integration domain, turns into
\begin{equation}\label{1}
\langle\sigma(t)\rangle=\epsilon
\int_0^{t}dx\psi_S(t-x)P_S(x)\int_x^{t}dt'\Psi_P(t').
\end{equation}
The long time limit shifts the main weight in the integral on the terms such that $x\lesssim t$.
Therefore  the approximation Eq. (\ref{p}), valid also for $P_S(t)$ in the long time limit, can be adopted.
Inserting such approximation in Eq. (\ref{1}) leads to:
\begin{equation}\label{2}
\langle\sigma(t)\rangle\simeq-\epsilon\frac{\sin\pi\mu_S}{\pi T_S^{\mu_S-1}}
\int_0^{t}dx\psi_S(t-x)x^{\mu_S-2}\int_x^{t}dt'\Psi_P(t').
\end{equation}
After getting rid of the integral in $t'$ by direct integration of  $\Psi_P(t)$,  one is left with  a simple  convolution which is easy to analyze in Laplace transform.
After some straightforward algebra, extracting the two asymptotic leading terms, leads to:
\BEQ
\langle \sigma(t)\rangle \simeq  k_1 t^{1-\mu_P} + k_2 t^{\mu_S-\mu_P-1}
\EEQ
 i.e. the asymptotic expression of Eq. (\ref{pi_din}), with the coefficients given by Eqs. (\ref{k1}) and (\ref{k2}).


\end{document}